\documentclass[aps,prl,amsmath,amssymb,showpacs,twocolumn,superscriptaddress]{revtex4-1}

\pdfoutput=1

\usepackage{graphicx}
\usepackage{dcolumn}
\usepackage{bm}
\usepackage{graphicx}
\usepackage{xspace}
\usepackage{multirow}
\usepackage{natbib}
\usepackage{color}
\usepackage{CJK}

\renewcommand{\vec}{\mathbf}

\begin{document}

\title{Angle-resolved photoemission studies of the superconducting gap symmetry in Fe-based superconductors}

\author{Y.-B. Huang} 
\affiliation{Beijing National Laboratory for Condensed Matter Physics, and Institute of Physics, Chinese Academy of Sciences, Beijing 100190, China}
\author{P. Richard} 
\affiliation{Beijing National Laboratory for Condensed Matter Physics, and Institute of Physics, Chinese Academy of Sciences, Beijing 100190, China}
\author{X.-P. Wang} 
\affiliation{Beijing National Laboratory for Condensed Matter Physics, and Institute of Physics, Chinese Academy of Sciences, Beijing 100190, China}
\author{T. Qian} 
\affiliation{Beijing National Laboratory for Condensed Matter Physics, and Institute of Physics, Chinese Academy of Sciences, Beijing 100190, China}
\author{H. Ding}\email{dingh@iphy.ac.cn}  
\affiliation{Beijing National Laboratory for Condensed Matter Physics, and Institute of Physics, Chinese Academy of Sciences, Beijing 100190, China}

\begin{abstract}
The superconducting gap is the fundamental parameter that characterizes the superconducting state, and its symmetry is a direct consequence of the mechanism responsible for Cooper pairing. Here we discuss about angle-resolved photoemission spectroscopy measurements of the superconducting gap in the Fe-based high-temperature superconductors. We show that the superconducting gap is Fermi surface dependent and nodeless with small anisotropy, or more precisely, a function of momentum. We show that while this observation is inconsistent with weak coupling approaches for superconductivity in these materials, it is well supported by strong coupling models and global superconducting gaps. We also suggest that the strong anisotropies measured by other probes sensitive to the residual density of states are not related to the pairing interaction itself, but rather emerge naturally from the smaller lifetime of the superconducting Cooper pairs that is a direct  consequence of the momentum dependent interband scattering inherent to these materials.   
\end{abstract}

\pacs{74.25.Jb, 74.70.Xa}



\maketitle

The superconducting (SC) gap is a direct fingerprint of the pairing mechanism in unconventional SC compounds such as the Fe-based high-temperature superconductors. Although numerous theoretical models have been proposed to account for experimental measurements, they can essentially be divided into two groups with seemingly orthogonal philosophies \cite{Richard_RoPP2011}: the ``weak coupling" and the ``strong coupling" approaches. According to the former type of approach, superconductivity is driven by interactions at the Fermi level ($E_F$), which may be related to spin fluctuations \cite{MazinPhysicaC2009,Graser_NJP2009,Hirschfeld_RoPP2011} as well as to orbital fluctuations \cite{KontaniPRL2010}. Within this framework, the SC gap of the Fe-based superconductors relies on the itinerancy of the electronic carriers and is mainly shaped by the Fermi surface (FS) topology and the properties in the vicinity of the FS, which varies from material to material. On the other hand, the SC pairing mechanism in the strong coupling approach is better described in the real space and the relevant energies are no longer limited to $E_F$. Even though the FS topology may play a role in determining the SC gap symmetry in the strong coupling approach as well \cite{HuJP_SR2012}, the dominant parameters are the local antiferromagnetic exchange interactions extracted from the magnetically-ordered parent compounds \cite{SeoPRL2008,C_Fang_PRX2011,YuR_arxiv2011,ZhouYi_EPL2011, HuJP_PRX2012}. An important consequence of this approach is that the SC gap symmetry is a property of the Brillouin zone (BZ) and the SC gap on a particular FS depends only on its absolute position in the momentum space. The mutual incompatibility of both kinds of approaches calls for a full experimental characterization of the SC gap. 

In contrast to the study of cuprates, the investigation of Fe-based superconductors reveals a fundamental and challenging complication: the latter materials are multi-band systems, and thus the SC gap needs to be determined on each FS separately, which requires a momentum-resolved probe. Angle-resolved photoemission spectroscopy (ARPES) is a powerful experimental technique used to observe directly the density of single-particle electronic excitations in the momentum space of crystals. In particular, the momentum-resolution capability of ARPES is a straightforward way to study the SC gap that opens at $E_F$ below the superconducting transition temperature ($T_c$). In this paper, we compare ARPES data of the SC gap obtained on various families of Fe-based superconductors. We show that the momentum distribution of the SC gap suggests that strong coupling approaches are more suitable to describe superconductivity in these materials.

\begin{figure*}[!t] 
\includegraphics[width=14cm]{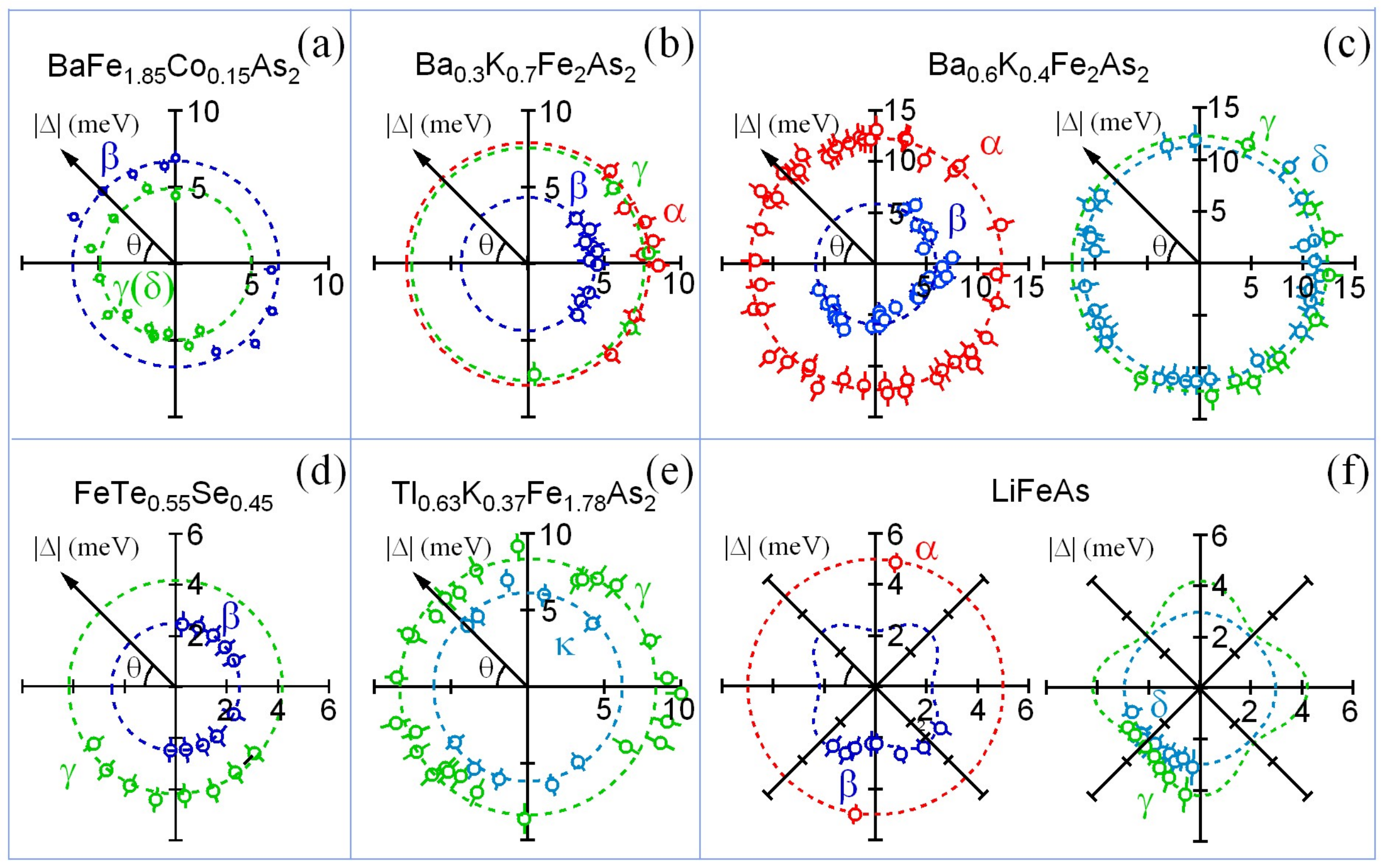} 
\caption{\label{Fig1_polar}(Color online) Polar distribution of the SC gap amplitude for various materials and FSs. The experimental data are extracted from previous experiments on (a) BaFe$_{1.85}$Co$_{0.15}$As$_2$ \cite{Terashima_PNAS2009}, (b) Ba$_{0.3}$K$_{0.7}$Fe$_2$As$_2$ \cite{Nakayama_PRB2011}, (c) Ba$_{0.6}$K$_{0.4}$Fe$_2$As$_2$ \cite{Nakayama_EPL2009}, (d) FeSe$_{0.55}$Te$_{0.45}$ \cite{MiaoPRB85}, (e) Tl$_{0.63}$K$_{0.37}$Fe$_2$Se$_2$ \cite{XP_WangEPL2011,XP_Wang_arxiv2012} and (f) LiFeAs \cite{UmezawaPRL108}. Each FS is labeled as in the paper from which the data have been extracted.} 
\end{figure*}

The first high-energy resolution ARPES studies of Ba$_{0.6}$K$_{0.4}$Fe$_2$As$_2$ reported FS dependent nodeless SC gaps with only little room for anisotropy \cite{Ding_EPL,L_Zhao_CPL2008}. Although quasi-nesting scattering was since then promoted to explain superconductivity because enhanced SC gaps were found only on quasi-nested FSs \cite{Ding_EPL}, the experimental observation of isotropic SC gaps poses a serious challenge to this same model since the FS topology and orbital distribution modulate the momentum dependence of the  electron-hole interband scattering believed to promote Cooper pairing. Upon hole or electron doping, the size of the hole and electron FS pockets evolve in opposite directions, with an immediate impact on the quasi-nesting conditions and the electron-hole interband scattering \cite{Ning_PRL2010}, thus providing critical opportunities to test the robustness of the SC gap isotropicity. Interestingly, nodeless and quite isotropic SC gaps were also observed by ARPES in underdoped Ba$_{0.75}$K$_{0.25}$Fe$_2$As$_2$ \cite{YM_Xu_UD}, overdoped Ba$_{0.3}$K$_{0.7}$Fe$_2$As$_2$ \cite{Nakayama_PRB2011} and electron-doped BaFe$_{1.85}$Co$_{0.15}$As$_2$ \cite{Terashima_PNAS2009}. Similarly, ARPES measurements indicate nodeless SC gaps as well in other materials with different FS topologies, crystal structures and cleaved surface terminations: Fe$_{1.07}$Se$_{0.3}$Te$_{0.7}$ \cite{Nakayama_PRL2010}, FeSe$_{0.55}$Te$_{0.45}$ \cite{MiaoPRB85}, NaFe$_{0.95}$Co$_{0.05}$As \cite{ZH_LiuPRB2011}, Tl$_{0.63}$K$_{0.37}$Fe$_2$Se$_2$\cite{XP_WangEPL2011,XP_Wang_arxiv2012}, A$_{0.8}$Fe$_2$Se$_2$ (A = K, Cs) \cite{Y_Zhang_NatureMat2011}, Tl$_{0.58}$Rb$_{0.42}$Fe$_{1.72}$Se$_2$ \cite{D_MouPRL2011} and NdFeAsO$_{0.9}$F$_{0.1}$ \cite{Kondo_PRL2008}. This impressive list of nodeless materials is in strong contradiction with the weak coupling approach, which can explain these results only by invoking a \emph{paradox} \cite{Hirschfeld_RoPP2011}. Yet, a close comparison between ARPES results and zero-field thermal conductivity rather suggests the reliability of the ARPES results \cite{Richard_RoPP2011}.

The robustness of the nodeless and almost isotropic SC gap of the Fe-based superconductors is well illustrated by the polar representation of the SC gap of various Fe-based superconductors given in Fig. \ref{Fig1_polar}. The only material so far for which moderate in-plane anisotropic, but yet nodeless, SC gaps have been detected by ARPES is LiFeAs \cite{UmezawaPRL108,BorisenkoSym2012}, for which the SC gap is displayed in Fig. \ref{Fig1_polar}(f). Although one could argue that this observation is compatible with orbital fluctuations \cite{BorisenkoSym2012}, we explain below that this result can be mainly explained by the strong coupling approach, which thus becomes a more likely candidate to unify the pairing mechanism in all the Fe-based materials listed above.

\begin{figure}[!t] 
\includegraphics[width=8.5cm]{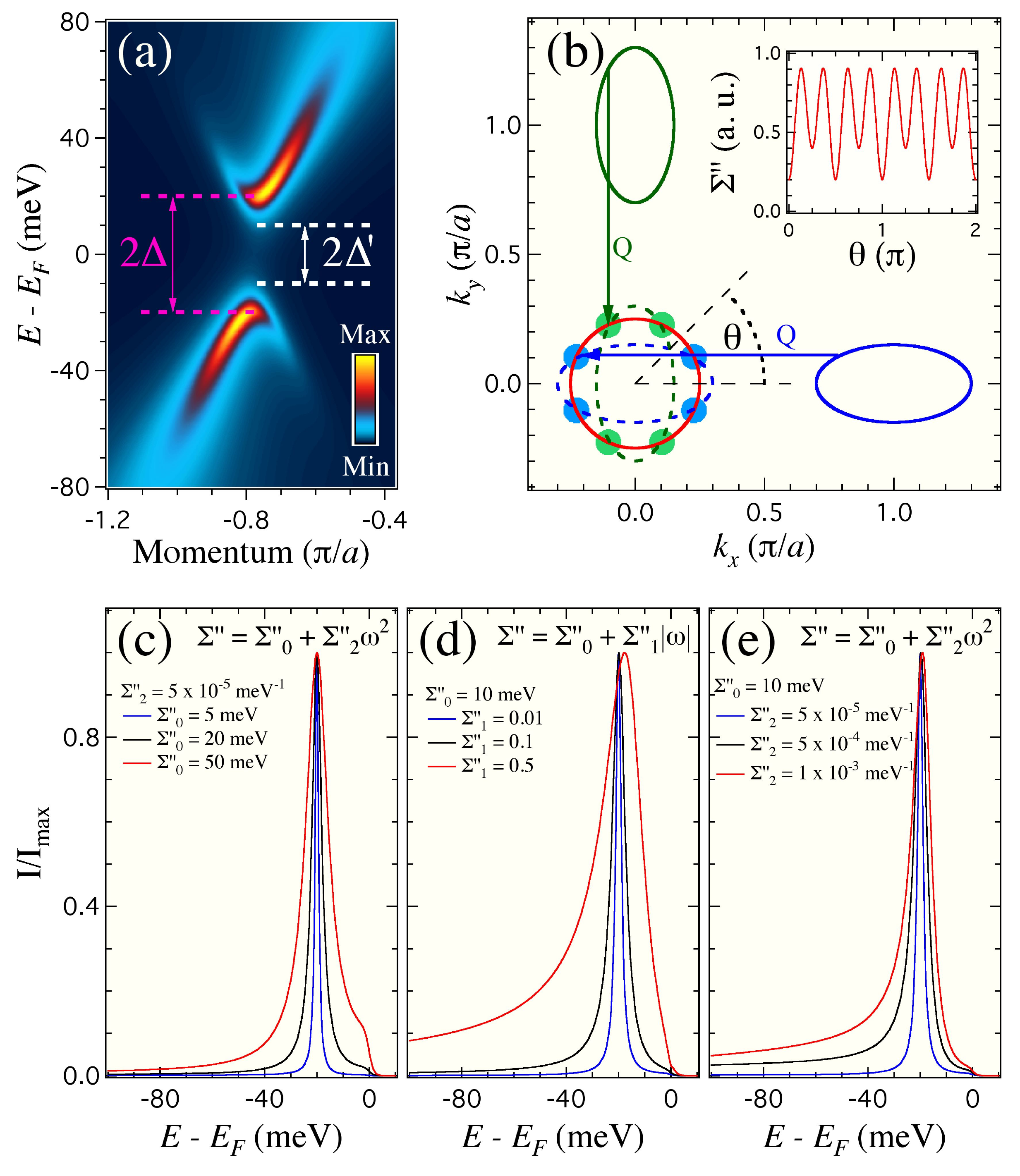} 
\caption{\label{Fig2_scattering}(Color online) (a) Simulation of the spectral function $A(\vec{k},\omega)$ in the presence of a 20 meV SC gap. We introduced an imaginary part to the self-energy with a quadratic dependence on energy in order to make the simulation more realistic. $\Delta$ corresponds to the SC gap while $\Delta^{\prime}$ is associated to an effective gap as would be measured by probes sensitive to a residual density of states. (b) Schematic FS of an hypothetical 2-band Fe-based superconductor. The dashed-line FSs have been translated by the antiferromagnetic wave vector Q to show where to expect stronger scattering (green and blue spots). The inset shows the schematic angular dependence of the imaginary part of the self-energy associated to interband scattering. (c)-(e) EDCs at $k_F$ for various parameters of the imaginary part of the self-energy $\Sigma^{\prime\prime}(\omega)=\Sigma^{\prime\prime}_0+\Sigma^{\prime\prime}_1|\omega|+\Sigma^{\prime\prime}_2\omega^2$ for the simulation in (a). The EDCs have be normalized by their maximum value.} 
\end{figure}

Despite the obvious trend in the ARPES measurements of observing isotropic or weakly anisotropic SC gaps, studies with alternative experimental techniques such as angle-resolved specific heat \cite{B_ZengNatCommun2010} rather reported strong gap anisotropies. What is responsible for such discrepancies? To answer this question, it is essential to first define clearly what we mean by SC gap. For this purpose, we plot in Fig. \ref{Fig2_scattering}(a) a simulation of ARPES data for a material in the SC state generated by using directly the BCS equations for the evolution of the band dispersion near $E_F$. We imposed $\Delta=20$ meV. For this demonstration, we removed the Fermi cutoff and introduced a non-zero scattering rate. As a result of particle-hole mixing, the simulation illustrates clearly the bending back of the electronic dispersion (Bogoliubov dispersion) at the Fermi wave vector ($k_F$). Albeit with finite energy and momentum resolutions, ARPES measures directly the band dispersion in the momentum space and it can therefore identify precisely the minimum gap location. In practice, as well as in theory, the size of the SC gap $\Delta$ characterizing the SC pairing interaction is simply given by the \emph{energy position of the band dispersion at the minimum gap location} (in other words the energy position of the ARPES SC coherent peak at the minimum gap location). In this particular case, ARPES will measure a SC gap $\Delta=20$ meV, the same value as we imposed. What would other probes measure? 

The answer to the previous question depends on which physical quantity is probed. In principle, spectroscopic tools sensitive to the electronic bands and the spectral function $A(\vec{k},\omega)$ (where $\omega=E-E_F$), like scanning tunneling spectroscopy (STS) or optical conductivity, for example, should observe the same value as ARPES. In contrast, some experimental techniques such as specific heat are rather very sensitive to \emph{non-zero density of states}. The size $\Delta^{\prime}$ of the gap measured in this case is thus determined by the energy for which the density of states becomes measurable rather than the position of the band dispersion at the minimum gap location. Should $\Delta$ and $\Delta^{\prime}$ have the same value? To answer this question, we first check the effect of different scattering on $A(\vec{k},\omega)$. Scattering affects directly the imaginary part $\Sigma^{\prime\prime}(\omega,\vec{k})$ of the self-energy $\Sigma(\omega,\vec{k})=\Sigma^{\prime}(\omega,\vec{k})+i\Sigma^{\prime\prime}(\omega,\vec{k})$. In Figs. \ref{Fig2_scattering}(c)-(e), we show the energy distribution curves (EDCs) at $k_F$ when using an imaginary part of the self-energy of the form $\Sigma^{\prime\prime}(\omega)=\Sigma^{\prime\prime}_0+\Sigma^{\prime\prime}_1|\omega|+\Sigma^{\prime\prime}_2\omega^2$, where $\Sigma^{\prime\prime}_0$, $\Sigma^{\prime\prime}_1$ and $\Sigma^{\prime\prime}_2$ are all positive constants. For sake of clarity, we first neglect the momentum dependence of scattering. Figs. \ref{Fig2_scattering}(c)-(e) illustrate clearly that one of the most drastic effect of increasing any of the previous parameters is to broaden the $k_F$ EDCs, as one can convince oneself by tracking the leading edge gap. As a corollary, the broadening induced by scattering creates a non-zero density of states inside the pairing gap $\Delta$. Any experimental probe sensitive to this residual density of states would thus measure an effective gap $\Delta^{\prime}$ smaller than $\Delta$.

Unlike conventional superconductors, for which the mean free path and the lifetime of the Cooper pairs are very long, the mean free path of Cooper pairs in high-temperature superconductors such as the Fe-based superconductors is quite small, and their lifetime is accordingly short. This means that scattering is indeed very important in the physics of these materials, and therefore it should not be a surprise that some experimental probes record gaps smaller than the SC gaps measured by ARPES. We now ask what can possibly act as a scatterer. Obviously, impurities can do so, and probably do so in these quite "dirty" materials where doping is introduced through chemical substitution, but the effect is expected to be momentum independent. Whether nearly-elastic interband scattering or quasi-nesting can lead to high-temperature superconductivity in the Fe-based superconductors is still a highly debated issue \cite{Richard_RoPP2011}. However, there is sufficient evidence showing that important interband scattering near $E_F$ occurs in these compounds. Indeed, bands quasi-nested by the antiferromagnetic wave vector $\vec{Q}$ show much broader SC coherent peaks in Ba$_{0.6}$K$_{0.4}$Fe$_2$As$_2$ \cite{Ding_EPL}, as well as a kink or anomaly in their electronic dispersion \cite{RichardPRL2009}. A pseudogap possibly originating from antiferromagnetic fluctuations also emerges in the underdoped regime of these materials \cite{YM_Xu_UD}. 

Unlike impurity scattering, interband scattering varies according to the FS topology and the momentum dependence of the corresponding self-energy can no longer be neglected. We illustrate this effect schematically in Fig. \ref{Fig2_scattering}(b) for the simplified case of a two-band system. Enhanced scattering on the central FS pocket occurs around the blue and green spots, which are connected to the FS pockets at (0,$\pi$) and ($\pi$,0) by $\vec{Q}$. As a function of the polar angle $\theta$, one should expect an antiferromagnetic scattering similar to the one described by the imaginary part of the self-energy displayed in inset, which has 8 minima. In general, the exact number and position of minima depends on the relative size and shape of the various FSs. For example, from the similar size of the FSs at $\Gamma$ and M in FeSe$_{0.55}$Te$_{0.45}$ \cite{MiaoPRB85}, one should expect only 4 minima in the scattering function, which is consistent with the 4 gap minima reported in a angle-resolved specific heat study \cite{B_ZengNatCommun2010}. The effective gap $\Delta^{\prime}$ measured by angle-resolved specific heat and other probes sensitive to residual density of states should be anisotropic in our simulation, even if the SC pairing gap itself is isotropic. Although $\Delta^{\prime}$ can serve as a good operational gap function for the fabrication of devices, it is obviously not appropriate to describe the pairing interaction due to its contamination from scattering. We also point out that this effect may be amplified under external magnetic field and when the gap size is relatively small, as in FeSe$_{0.55}$Te$_{0.45}$ \cite{MiaoPRB85}.

\begin{figure}[!t] 
\includegraphics[width=8.5cm]{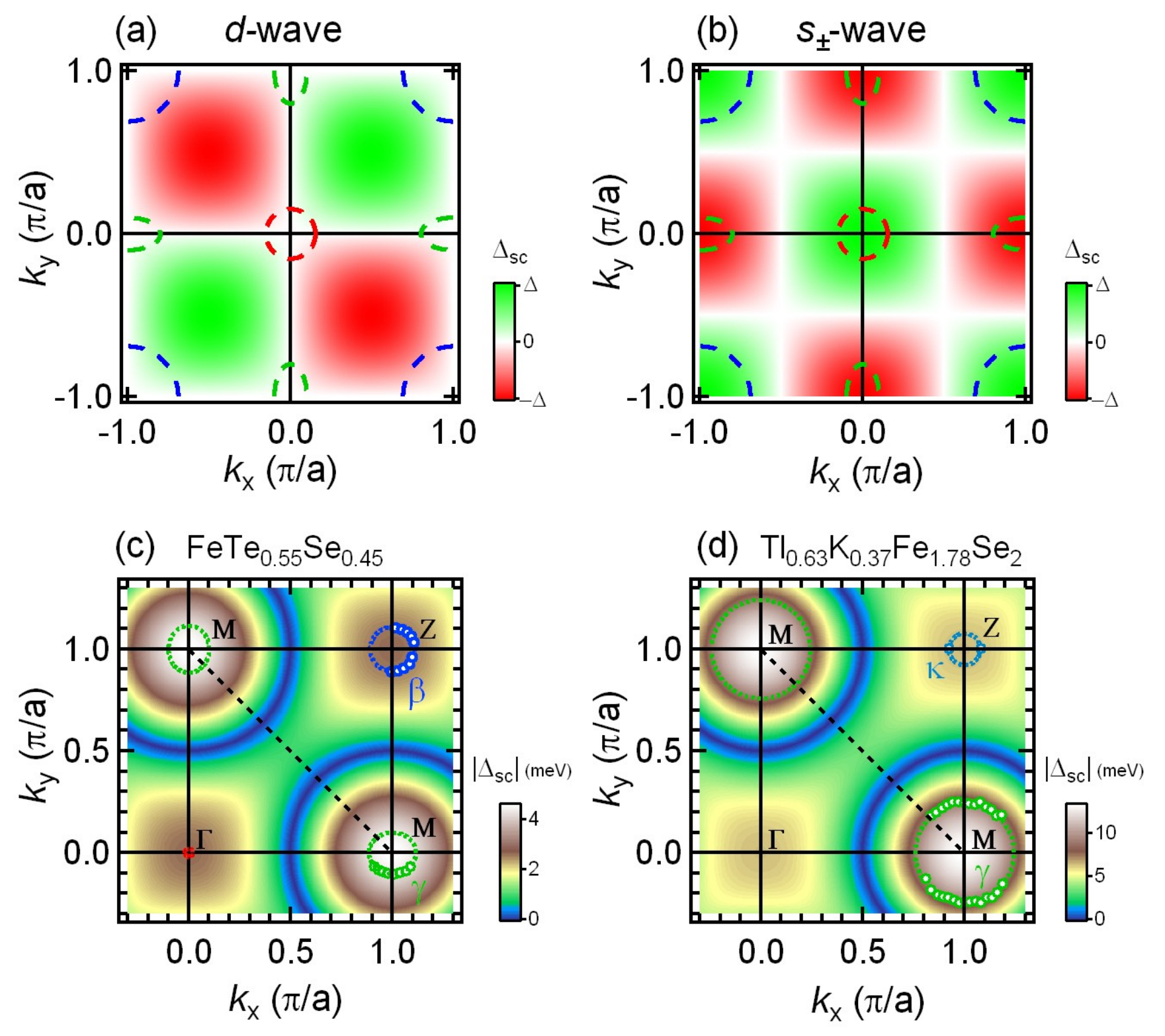}
\caption{\label{Fig3_Jmap}(Color online) (a) Intensity of the global SC gap function $\Delta_d=\Delta_2\sin k_x \sin k_y$ obtained by considering the next-nearest neighbor exchange coupling ($J_2$) in the $d$-wave channel. The schematic FSs of a typical ferropnictide are overlapped. (b) Same as (a) but for the $\Delta_s\pm=\Delta_2 \cos k_x \cos k_y$ function derived in the $s_{\pm}$-wave channel. (c) Absolute value of the global SC gap function and FSs reported for FeSe$_{0.55}$Te$_{0.45}$, which takes the form $|\Delta_2 \cos k_x \cos k_y + (\Delta_3/2)(\cos 2k_x+\cos 2k_y)|$ \cite{MiaoPRB85}. (d) Same as (c) but for Tl$_{0.63}$K$_{0.37}$Fe$_2$Se$_2$ \cite{XP_WangEPL2011,XP_Wang_arxiv2012}.} 
\end{figure}

We now describe the notion of global gap that is central to the strong coupling approach. Within this approach, the magnetic interactions are simply described in the real space using $\delta (\vec{r_{i}}-\vec{r_{j}})$ functions at the distance between the neighbors $i$ and $j$ considered. Their momentum dependence, obtained by performing a Fourier transformation, are then expressed as a combination of simple sine and cosine functions \cite{HuJP_SR2012}. The global SC gap determined from the strong coupling approach are naturally proportional to these functions. For example, the next-nearest neighbor interactions in the $d$-wave and $s$-wave channels lead to SC gaps of the forms $\Delta_d=\Delta_2\sin k_x \sin k_y$ and $\Delta_s\pm=\Delta_2 \cos k_x \cos k_y$, as illustrated in Figs. \ref{Fig3_Jmap}(a) and \ref{Fig3_Jmap}(b), respectively. In the same figures, we trace the schematic FS of the ferropnictide superconductors, for which the next-nearest neighbor antiferromagnetic exchange coupling constant ($J_2$) dominates \cite{HuJP_SR2012}. The SC gap expected for each $k_F$ value depends simply on its position $(k_x,k_y)$. Figure \ref{Fig3_Jmap}(a) indicates clearly that the $d$-wave pairing is not favorable for this material since each FS falls at momentum locations for which the global SC gap is very small. In contrast, the $s_{\pm}$-wave function offers a good optimization of the SC gap on each FS. 

Sometimes, magnetic interactions between two sites are not sufficient to describe the magnetic ground states of the studied materials, and more parameters are needed. For example, inelastic neutron scattering experiments suggest that unlike the ferropnictides, the interaction between next-next-nearest neighbors is not negligible in the ferrochalcogenides \cite{LipscombePRL2011}. This experimental observation is translated in the momentum space by the introduction of a function of the form $(\Delta_3/2)(\cos 2k_x+\cos 2k_y)$ for the SC gap function. In Figs. \ref{Fig3_Jmap}(c) and \ref{Fig3_Jmap}(d), we overlap the FSs and the absolute value of the global SC gap functions derived experimentally for FeSe$_{0.55}$Te$_{0.45}$ \cite{MiaoPRB85} and Tl$_{0.63}$K$_{0.37}$Fe$_2$Se$_2$ \cite{XP_WangEPL2011,XP_Wang_arxiv2012}, respectively. For FeSe$_{0.55}$Te$_{0.45}$, the SC gap is larger at the M point \cite{MiaoPRB85}, in agreement with the strong coupling approach. More importantly, while the observation of large SC gap in the absence of hole FS at the $\Gamma$ point in the 122-ferrochalcogenides illustrates the failure of the electron-hole quasi-nesting scenario to explain superconductivity in this family of compounds \cite{XP_WangEPL2011,XP_Wang_arxiv2012,Y_Zhang_NatureMat2011,D_MouPRL2011,Qian_PRL2011}, the presence or not of FS at $\Gamma$ is totally less relevant in the framework of the strong coupling approach. Indeed, the global gap structure of Tl$_{0.63}$K$_{0.37}$Fe$_2$Se$_2$, which can be viewed as an (Tl, K)-intercalated 11-chalcogenide, has the same form as in FeSe$_{0.55}$Te$_{0.45}$.

\begin{figure}[!t] 
\includegraphics[width=8.5cm]{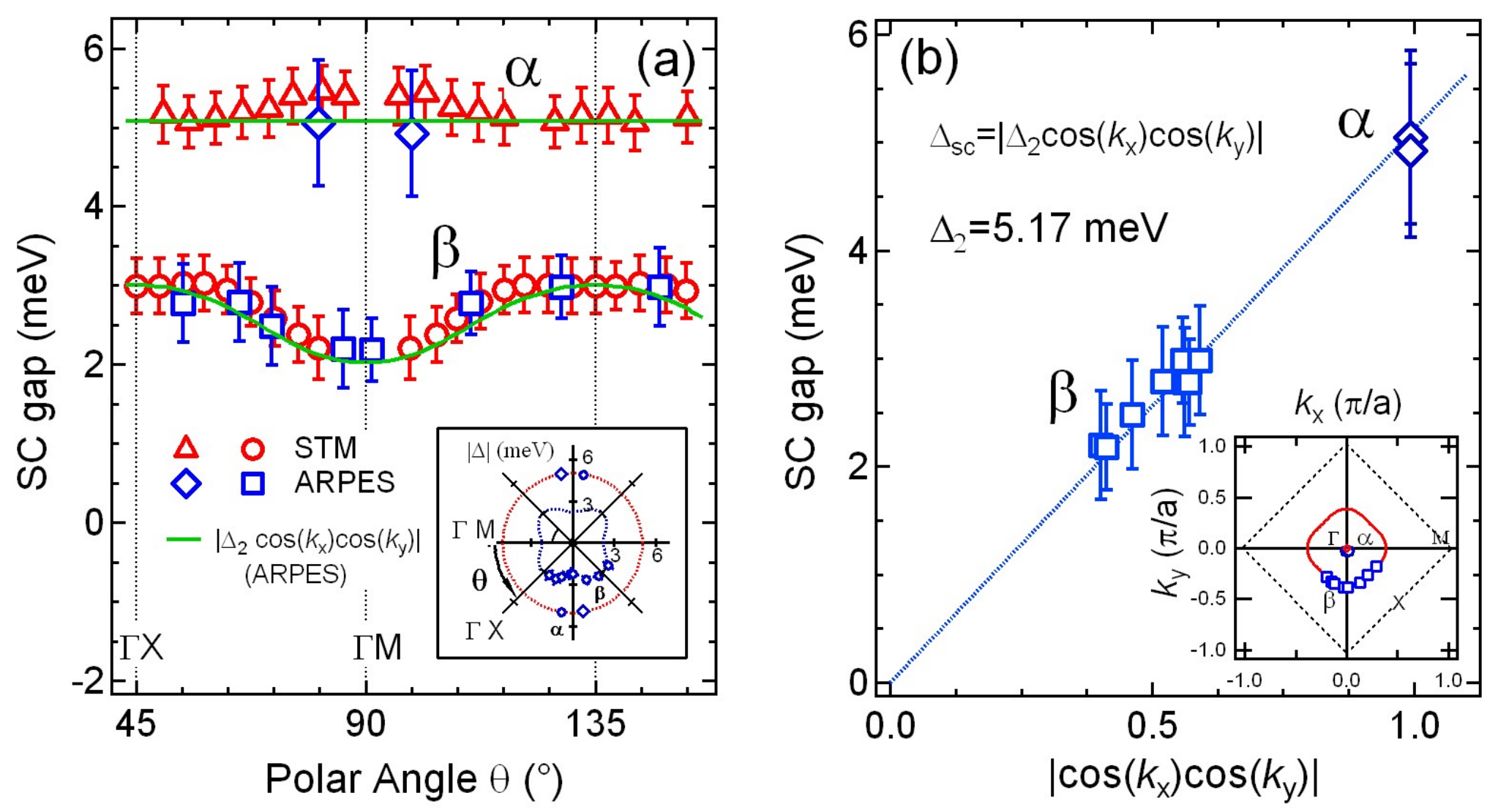} 
\caption{\label{Fig4_STM}(Color online) (a) Comparison of the polar dependence of the SC gap in LiFeAs measured by ARPES \cite{UmezawaPRL108} and Fourier transform STS \cite{Allan_Science336}. The green curves represent the fit of the ARPES data to the strong coupling function $|\Delta_2\cos k_x\cos k_y|$. The insets shows a polar representation of the same ARPES data. (b) Fit of the ARPES data on the strong coupling function $|\Delta_2\cos k_x\cos k_y|$ using a single global parameter. The inset indicates the momentum locations where the superconducting gap was recorded. } 
\end{figure}

\begin{figure*}[!t] 
\includegraphics[width=14cm]{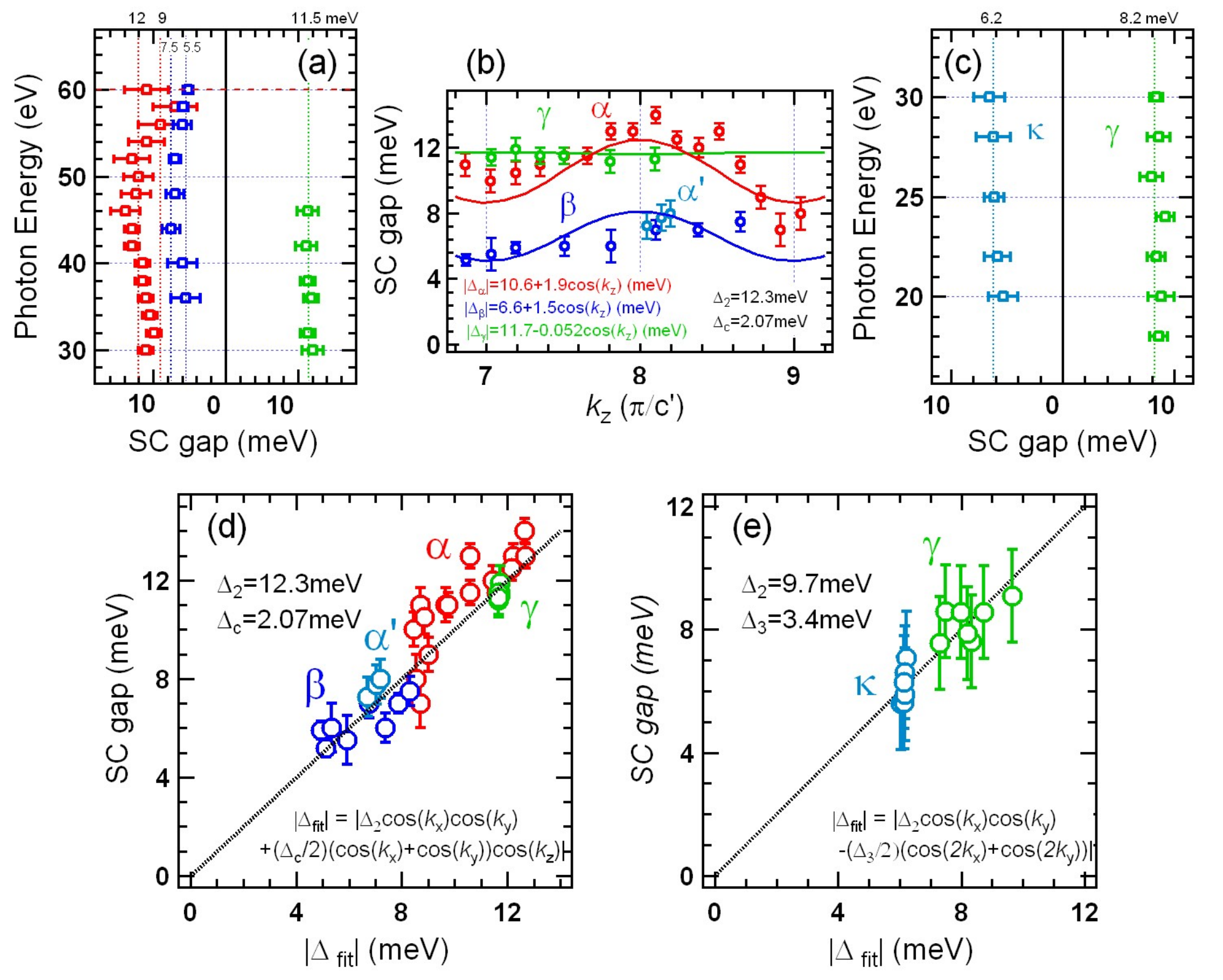} 
\caption{\label{Fig5_fit}(Color online) (a) ARPES measurement of the SC gap on various FS sheets in Ba$_{0.6}$K$_{0.4}$Fe$_2$As$_2$ as a function of photon energy \cite{YM_Xu_NPhys2011}. (b) Same data but plotted as a function of the perpendicular momentum $k_z$ \cite{YM_Xu_NPhys2011}. The solid curves illustrate the $k_z$ dependence of the SC gap. (c) Same as (a) but for Tl$_{0.63}$K$_{0.37}$Fe$_2$Se$_2$. The data are extracted from refs. \cite{XP_WangEPL2011,XP_Wang_arxiv2012}. (d) Fit of the ARPES gap data on Ba$_{0.6}$K$_{0.4}$Fe$_2$As$_2$ \cite{YM_Xu_NPhys2011} on a global SC gap function that includes an inter-plane coupling. (e) Fit of the ARPES gap data on Tl$_{0.63}$K$_{0.37}$Fe$_2$Se$_2$ \cite{XP_WangEPL2011,XP_Wang_arxiv2012} on a 2D global SC gap function.} 
\end{figure*}

The intensity plots shown in Figs. \ref{Fig3_Jmap}(b)-\ref{Fig3_Jmap}(d) suggest that the SC gaps measured in the ferropnictides and ferrochalcogenides should be more or less isotropic. For example, a fit of the strong coupling function to experimental results on NaFe$_{0.95}$Co$_{0.05}$As indicate only very small anisotropy \cite{ZH_LiuPRB2011}. However, the strong coupling approach does not prevent some anisotropies for certain shapes of FS. In Fig. \ref{Fig4_STM}(a), we compare the angular dependence of the SC gap derived from ARPES \cite{UmezawaPRL108} and Fourier transform STS \cite{Allan_Science336} measurements of LiFeAs. Both series of results agree remarkably well, indicating that they describe exactly the same physics. Although the $\alpha$ band exhibits an almost constant SC gap, the gap size on the $\beta$ FS varies from 2 to 3 meV, which is not negligible. As shown in the inset of Fig. \ref{Fig4_STM}(a), a $\cos 4\theta$ term is necessary to expressed the polar dependence of the SC gap on the $\beta$ FS \emph{taken separately} \cite{UmezawaPRL108,BorisenkoSym2012}. This representation becomes even simpler when we consider a global SC gap function. In Fig. \ref{Fig4_STM}(b), we show a fit of the $\beta$ and $\alpha$ bands with a single global parameter. Interestingly, the SC gap for both bands fall perfectly on the $\cos k_x\cos k_y$ gap function, showing that the observed anisotropy is a mere consequence of the shape of the FSs. We caution though that an anisotropic behavior that cannot simply described in terms of a global SC gap is also observed around the M point \cite{UmezawaPRL108,BorisenkoSym2012}, as shown in Fig. \ref{Fig1_polar}(f). However, this behavior might be explained in terms of band hybridization \cite{UmezawaPRL108}. 

So far we considered only 2D electronic band structures. However, the band structure of the Fe-based superconductors has a non-negligible 3D component \cite{Richard_RoPP2011} that should be included for a complete characterization of the SC gap. In practice, ARPES can access the band dispersion perpendicularly to the probed surface by varying the energy of the incident photons \cite{DamascelliPScrypta2004}. We show in Fig. \ref{Fig5_fit}(a) the results obtained on Ba$_{0.6}$K$_{0.4}$Fe$_2$As$_2$ over a wide photon energy range \cite{YM_Xu_NPhys2011}. While the SC gap on the $\beta$ and $\gamma$ FSs are almost constant, the amplitude of the SC gap on the $\alpha$ band is clearly modulated as photon energy is tuned. This effect can be better expressed by plotting the data with respect to the perpendicular momentum $k_z$ \cite{YM_Xu_NPhys2011}, as illustrated in Fig. \ref{Fig5_fit}(b). In contrast to Ba$_{0.6}$K$_{0.4}$Fe$_2$As$_2$, Fig. \ref{Fig5_fit}(c) indicates that all the SC gap in Tl$_{0.63}$K$_{0.37}$Fe$_2$Se$_2$ are nearly independent of $k_z$ \cite{XP_Wang_arxiv2012}. It is interesting to point out though that none of the hole FS pockets, on which $k_z$ variation of the gap is observe in Ba$_{0.6}$K$_{0.4}$Fe$_2$As$_2$, emerges at $E_F$ in Tl$_{0.63}$K$_{0.37}$Fe$_2$Se$_2$.  

Figure \ref{Fig5_fit}(d) shows explicitly the fit of the various FSs in Ba$_{0.6}$K$_{0.4}$Fe$_2$As$_2$ on a single global gap function with only two parameters, that is to say $\Delta_2$ associated to the local exchange coupling constant $J_2$ prevailing in this material, and $\Delta_c$ characterizing the inter-plane coupling. Despite little deviation, the fit shows a reasonable agreement with the experimental data. Similarly, Fig. \ref{Fig5_fit}(e) also shows explicitly the fit of the various FSs in Tl$_{0.63}$K$_{0.37}$Fe$_2$Se$_2$ on a single global gap function with only two parameters. It this particular case though, the inter-plane coupling is neglected but the gap function includes the contributions from both $J_2$ and $J_3$, the non-negligible next-next-nearest neighbor exchange coupling constant in the ferrochalcogenides \cite{XP_Wang_arxiv2012}. Within error bars, the gap function fits the data pretty well. 

In searching for the mechanism leading to high-temperature superconductivity, simplicity and universality appear as fundamental criteria. Not only the global gap functions derived from the strong coupling approach can have their form predicted from the analysis of independent inelastic neutron scattering measurements of spin-wave dispersions, they are indeed very simple, with only a few parameters. The remarkable agreement between the ARPES gap data and the inelastic neutron scattering data via the gap functions derived from the strong coupling models suggest that the physics contained in these models is at least partly appropriate to describe the Fe-based superconductors. More importantly, not only such approach seems valid for all Fe-based superconductors, as we showed in this paper, it is also consistent with the gap symmetry of the cuprates \cite{HuJP_SR2012}. 

In summary, we presented sufficient evidences to show that the weak coupling approaches are most likely inadequate to describe properly the SC gap in the Fe-based superconductors. The most significant are the quite robust nodelessness of the SC gap, which is free of strong anisotropy in almost every Fe-based superconductors, and the absence of hole FS pocket in the 122-ferrochalcogenides, thus preventing any electron-hole FS quasi-nesting. In contrast, the use of global functions as derived from the strong coupling approach can explained the relative lack of anisotropy in the SC gap of these systems. In addition, it can explain the stronger anisotropy detected in LiFeAs. Although the current models may still need refinement, our ARPES measurements clearly suggest that the strong coupling approach is more suitable for a universal description of superconductivity in the Fe-based superconductors.

\section*{Acknowledgements}

We are grateful to J. C. Davis, H.-H. Wen, J.-P. Hu, Z. Fang and X. Dai for useful discussions. This work is supported by the Chinese Academy of Sciences (grant No. 2010Y1JB6), the Ministry of Science and Technology of China (grants No. 2010CB923000, No. 2011CBA0010) and the Nature Science Foundation of China (grants No. 10974175, No. 11004232, and No. 11050110422).

\bibliography{biblio_en}

\end{document}